# Effect of La Doping on Microstructure and Critical Current Density of MgB$_2$

*Chandra Shekhar[a], Rajiv Giri[a], R S Tiwari[a], D S Rana[b], S K Malik[b] and O N Srivastava[a]**

[a] Department of Physics, Banaras Hindu University, Varanasi 221005, India
[b] Tata Institute of Fundamental Research, Mumbai-400005, India

## Abstract

In the present study, La-doped MgB$_2$ superconductors with different doping level (Mg$_{1-x}$La$_x$B$_2$; x=0.00, 0.01, 0.03 & 0.05) have been synthesized by solid-state reaction route at ambient pressure. Effect of La doping have been investigated in relation to microstructural characteristics and superconducting properties, particularly intragrain critical current density (J$_c$). The microstructural characteristics of the as synthesized Mg(La)B$_2$ compounds were studied employing transmission electron microscopic (TEM) technique. The TEM investigations reveal inclusion of LaB$_6$ nanoparticles within the MgB$_2$ grains which provide effective flux pinning centres. The evaluation of intragrain J$_c$ through magnetic measurements on the fine powdered version of the as synthesized samples reveal that J$_c$ of the samples change significantly with the doping level. The optimum result on J$_c$ is obtained for Mg$_{0.97}$La$_{0.03}$B$_2$ at 5K, the J$_c$ reaches ~1.4x10$^7$A/cm$^2$ in self field, ~2.1 x 10$^6$A/cm$^2$ at 1T, ~2.5 x 10$^5$A/cm$^2$ at 2.5T and ~1.8 x 10$^4$ A/cm$^2$ at 4.5T. The highest value of intragrain J$_c$ in Mg$_{0.97}$La$_{0.03}$B$_2$ superconductor has been attributed to the inclusion of LaB$_6$ nanoparticles which are capable of providing effective flux pinning centres.

*Corresponding author

E-mail: hepons@yahoo.com, girirajiv_bhu@yahoo.com

**Introduction**

The discovery of superconductivity at 40K in $MgB_2$ by Nagamatsu et al. [1] has generated a great deal of excitement in both fundamental and practical investigations of this material. One advantage of $MgB_2$ is in regard to its applications at comparatively higher temperatures (20-30K) region, where the conventional superconductors can not operate. Also, the progress in cryogen free cooling technique at temperature region of 20-30K will promote the development and application of $MgB_2$. Another important aspect in $MgB_2$ bulk material is that supercurrent flow is largely unhindered by grain boundaries (GBs) [2,3], thus providing a high feasibility of scaling up the material to form bulk shapes like wire and tapes. For widespread applications of $MgB_2$ high value of critical current density in higher magnetic field is required and hence need of adequate flux pinning centres is evident. The effective flux pinning is known to occur when core of fluxoids find higher density normal materials to sit on. Strong pinning centres prevent flux melting and the consequent creep. If local structural perturbations are strong and these are isolated, they can act as effective flux pinning centres provided that their sizes are compatible with the coherence length (~6nm). Unfortunately, the bulk materials prepared at ambient pressure always show lower $J_c$ value due to poor grain connectivity, low density and poor flux pinning [4,5]. It is found that most efficient techniques used to fabricate high quality $MgB_2$ samples are hot-isostatic pressing and uniaxial hot pressing at high pressures [6,7]. Also, proton and neutron irradiation are applied to enhance $J_c$ properties in $MgB_2$ superconductors [8,9]. However, these techniques are not suitable for the preparation of $MgB_2$ wire and tapes because of their complicated nature and the practical difficulty in the use of very large vessels and dyes in case of high temperature and high pressure. On the other hand, it is reported that chemical doping is effective to increase $J_c$ in high-$T_c$ cuprates superconductors [10]. Similar techniques have been also utilized for the $MgB_2$ materials. Recently Dou et al. [11] and Matsumoto et al. [12] have synthesized $MgB_2$ superconductor in bulk form at ambient pressure through doping of SiC and $SiO_2$ & SiC respectively. They have reported enhancement of flux pinning properties due to nano inclusion of SiC (10-100 nm) into $MgB_2$ grains. Many groups have been trying to improve the $J_c$ by chemical doping. At 20K, pure $MgB_2$ generally has a intragrain $J_c$ of about $10^4$ $A/cm^2$ under 1T and irreversible field ($H_{irr}$) is about 4T [7, 13,14]. It may be pointed out that the estimation of $J_c \sim 10^4$ $A/cm^2$ (at 1T and 20K) is for undoped samples devoid of any flux pinning centers. However, when doping is done to produce flux pinning centers, this estimate generally become $\sim 10^5$ $A/cm^2$ (at 1T and 20K) [15-18]. The Ti doping (few to ~10 atomic%) in $MgB_2$ has been known to enhance intragrain $J_c$ (evaluated through magnetization measurement and using Bean's formula) values upto $10^5$ $A/cm^2$ under 1T and $H_{irr}$ near 5T at 20K with small $T_c$ reduction [15, 16]. Shu-Fang Wang et al. [17] have reported that $J_c$ increases up to 5 x $10^5$ $A/cm^2$ at 5K and in self field for Cd doped $MgB_2$. Rui et al. have reported enhancement of $J_c$ by adding nano-alumina (10-50nm) into $MgB_2$ [18]. It should be pointed out that the above authors have used Bean's formula to estimate intragrain $J_c$. Recently doping of Al, Fe, Au, Pb into $MgB_2$ have been reported for improvement of flux pinning centres and consequently enhancement in intragrain $J_c$ values [19-22]. To date, two important factors of low density and poor flux pinning are main obstacles to obtain high $J_c$ value in $MgB_2$ superconductor. Therefore, it is necessary to carry out further research by addition and doping of suitable elements into $MgB_2$ to understand the mechanism of the flux pinning in $MgB_2$ and eventually realize the widespread applications of $MgB_2$. It is known that microstructural features affect crucially the critical current density of superconducting materials. Therefore, in order to explore the microstructural characteristics and its possible correlation with superconducting properties, particularly $J_c$, in the present paper, we have studied structural, microstructural characteristics of the as synthesized $Mg_{1-x}La_xB_2$ ($0.00 \leq x \leq 0.1$) samples employing transmission electron microscopic (TEM) technique. TEM investigation reveals nanoparticle $LaB_6$ inclusions within the $MgB_2$ grains.

The intragrain critical current density ($J_c$) evaluated through magnetic measurement for sample with various compositions have been found to vary significantly e.g. $J_c$ of $Mg_{0.97}La_{0.3}B_2$ sample is ~2.1 x $10^6$ $A/cm^2$ and for $Mg_{0.99}La_{0.01}B_2$ sample, it is ~6.5x$10^5$$A/cm^2$ at 5K and 1T. A correlation between intragrain $J_c$ and details of the microstructure has been shown to exist. The $Mg_{0.97}La_{0.3}B_2$ sample which possesses the highest intragrain $J_c$ ~1.4x$10^7$$A/cm^2$ in self field at 5K exhibits $LaB_6$ nanoparticle in the $MgB_2$ grains.

**Experimental details**

The synthesis of La doped $MgB_2$ with nominal composition of $Mg_{1-x}La_xB_2$ ($0.00 \leq x \leq 0.1$) has been carried out by solid state reaction method at ambient pressure by employing a special encapsulation technique developed in our laboratory. The powders [Mg (99.9%), La (99%) and B (99%)] were fully mixed and were cold pressed (3.0 tons/inch$^2$) into small rectangular pellets (10 x 5 x 1) mm$^3$. Then the pellets of $Mg(La)B_2$ were encapsulated in a Mg metal cover to circumvent the formation of MgO during sintering process. The pellet configuration was put on a Ta boat and sintered in flowing Ar atmosphere in a tube furnace at 600°C for 1h, at 800°C for 1h and at 900°C for 2h. The pellet was cooled to room temperature at the rate of 100°C/h. The pellet was taken out and encapsulating Mg cover was removed. The details of the synthesis were similar as described in our earlier publication [23]. All the samples in the present investigation were subjected to gross structural characterization by X-ray diffraction technique (XRD, Philips PW-1710 CuK$_\alpha$), electrical transport characterization by four-probe technique (Keithley resistivity Hall set-up), the microstructural characterization by transmission electron microscope (Philips EM-CM-12) and elemental composition was determined by energy dispersive X-ray analysis (EDAX, Oxford-ENCA). The magnetization measurements have been carried out at Tata Institute of Fundamental Research (Mumbai, India) over a temperature range of 5-40K employing a physical property measurement system (PPMS, Quantum Design). Intragrain $J_c$ (magnetic $J_c$) was calculated from the height '$\Delta$M' of the magnetization loop (M-H) using Bean's critical state model [24]. It should be pointed out that Bean's formula leads to the optimum estimate of intragrain $J_c$ for superconductors having weakly coupled grains. It is not quite appropriate for bulk $MgB_2$ superconductors because of strong grain coupling in this material. The magnetization measurements have been carried out on fine ground powders of the as synthesized samples. In the fine powder form, strong coupling is non-existent, the intragrain critical current density can be estimated employing Bean's formula and using average size of the powder particles. Usually, the particles after grinding of samples may not correspond to singular grains but are as estimated through SEM small agglomerates of nearly spherical shape ( ~5 μm ) covering only few grains.  The intragrain $J_c$, therefore, can be estimated by the following formula :

$$J_c = \frac{30\Delta M}{<d>}$$

Where '$\Delta$M' is change in magnetization with increasing and decreasing field (in emu/cm$^3$) and 'd' is average particle size (in cm).

**Results and discussion**

The dc magnetic susceptibility ($\chi$) of $Mg_{1-x}La_xB_2$ (with x=0.01, 0.03, 0.05) superconducting samples are shown in Fig. 1 for 50 Oe field as a function of temperature. The $\chi$-T curve shows diamagnetic onset transition temperature for $Mg_{0.99}La_{0.01}B_2$, $Mg_{0.97}La_{0.03}B_2$ and $Mg_{0.95}La_{0.05}B_2$ samples to be 40K, 39K and 37K respectively. The central aim of the present investigation is to explore the flux pinning centres originating as a result of doping. We first describe various microstructural features induced by different doping levels of La at Mg site of $MgB_2$ compound. Thereafter, evaluation of critical current density through magnetic hysteresis loop and correlation between microstructural features and intragrain $J_c$ will be described and discussed. In order to investigate microstructural features of La doped $MgB_2$, we carried out extensive studies of nature of grain boundaries (GBs) and inclusion of secondary particles by employing the technique of TEM, which is considered to be the viable technique for such studies.

Fig. 2(a) shows a representative transmission electron micrograph for $Mg_{0.99}La_{0.01}B_2$ compound. The dominant feature resulting from this doping is the occurrence of $LaB_6$ nanoparticles which are found to be invariably present. The selected area diffraction pattern corresponding to TEM micrograph is shown in Fig. 2(b). The slight splitting of diffraction spots reveals the presence of low angle grain boundaries in $Mg_{0.99}La_{0.01}B_2$ compound.

Fig. 2(c) represents the TEM micrograph of $Mg_{0.97}La_{0.03}B_2$ compound. The dominant and specific microstructural feature for this compound is the presence of high density of $LaB_6$ nanoparticles within $MgB_2$ grains. The size of these nanoparticles lies in the range of 5-15nm. Among these nano inclusions, those which are compatible with coherence length of $MgB_2$ (~6nm) may work as effective flux pinning centres. The SAD pattern corresponding to TEM micrograph [shown in Fig. 2(d)] reveals the spotty ring pattern. These diffraction rings, which correspond to $LaB_6$ and $MgB_2$, depict the inclusion of $LaB_6$ nanoparticle in $MgB_2$ grains.

The TEM micrograph for $Mg_{0.95}La_{0.05}B_2$ revealing the presence of $LaB_6$ nanoparticles (20-35 nm) is discernible from the micrograph Fig. 2(e). It is interesting to note that the size of $LaB_6$ nanoparticle in this compound is bigger than those of the $Mg_{0.97}La_{0.03}B_2$ compound [as shown in Fig. 2(c)] and has low density of nanoparticles. The SAD pattern corresponding to TEM micrograph [shown in fig. 2(e)] reveals hexagonal arrangement of diffraction spots of $MgB_2$ along with square diffraction spots of cubic $LaB_6$. The diffraction spots of the SAD pattern of $Mg_{0.95}La_{0.05}B_2$ [Fig. 2(f)] have been indexed. These results clearly show that with different doping concentration of La at Mg site of $MgB_2$ their microstructural features manifested by the occurrence of $LaB_6$ nanoparticles vary significantly.

The magnetization measurements as a function of magnetic fields have been carried out at temperature 5K, 10K, 20K and 30K, for each samples in the powder form. It may further be pointed out that several workers have employed Bean's formula for the undoped and doped $MgB_2$ samples [7, 21, 25, 26].

Keeping the above in view, in the following we will proceed to describe the estimation of $J_c$ (intragrain) for the present $Mg_{1-x}La_xB_2$ samples as obtained on powder form. The intragrain $J_c$ as function of magnetic field at temperatures of 5K, 10K, 20K and 30K for $MgB_2$, $Mg_{0.99}La_{0.01}B_2$, $Mg_{0.97}La_{0.03}B_2$ and $Mg_{0.95}La_{0.05}B_2$ are shown in Fig. 3(a), 3(b), 3(c) and 3(d) respectively. It is clear from $J_c$ vs H curves, the intragrain $J_c$ of $Mg_{0.97}La_{0.03}B_2$ compound attains the highest value among all the compounds for all temperature and the whole field region upto 5T. This compound contains high density of $LaB_6$ nanoparticles (~5-15 nm) inclusion into $MgB_2$ matrix. As for example at 5K, the $J_c$ of $Mg_{0.97}La_{0.03}B_2$ compound is ~1.4 x $10^7$A/cm$^2$ in self field and ~2.1 x $10^6$/cm$^2$ at 1T, ~2.5x $10^5$A/cm$^2$ at 2.5T and ~1.8 x $10^4$A/cm$^2$ at 4.5T. For $Mg_{0.99}La_{0.01}B_2$ compound, which is nearly without inclusion of $LaB_6$ nanoparticles, $J_c$ values at 5K are ~6.0 x $10^6$A/cm$^2$ in self field, ~6.5 x $10^5$A/cm$^2$ at 1T, ~8.6 x $10^4$A/cm$^2$ at 2.5T and ~5.6 x $10^3$A/cm$^2$ at 4.5T. The intragrain $J_c$ values for $Mg_{0.95}La_{0.05}B_2$ compound, having low density of larger $LaB_6$ nanoparticles in comparison to $Mg_{0.97}La_{0.03}B_2$ compound, are also lower than $Mg_{0.97}La_{0.03}B_2$. The $J_c$ value for $Mg_{0.95}La_{0.05}B_2$ compound achieves the value of ~1.2x$10^6$A/cm$^2$, ~2.6 x $10^5$A/cm$^2$, ~4.4 x $10^4$A/cm$^2$ and ~1.0 x $10^4$A/cm$^2$ at 5K in self field, 1T, 2.5T and 4.5T respectively. Similar variations of $J_c$ with magnetic field were also observed at temperatures 10K, 20K and 30K. Table-1 brings out the comparision of intragrain $J_c$ values for the undoped and optimally doped $MgB_2$ samples in the self field and various magnetic fields. As can be seen from this table, Jcs for La doped $MgB_2$ sample ($Mg_{0.97}La_{0.03}B_2$) is higher than undoped version for all fields.

These results are first of their type for $MgB_2$ superconductors. All the $J_c$ data reported here by us are much higher than the best results reported so far for $MgB_2$ bulk materials including those prepared under high pressure (the typical value of $J_c$ is ~2 x $10^4$ A/cm$^2$ at 20K and 1T) [7], Ti doped $MgB_2$ (the typical value of $J_c$ is ~2 x $10^6$ A/cm$^2$ at 5K and in self field) [15], Cd doped $MgB_2$ (the typical value of $J_c$ at 5K and in self field is ~5 x $10^5$ A/cm$^2$) [17] and Au coated $MgB_2$ thin film (the typical value of $J_c$ is ~1.22 x $10^7$ A/cm$^2$ at 5K and in self field [21].

It may be noticed from the $J_c$ vs H behavior of La doped $MgB_2$ samples that $J_c$ decreases slowly with increasing magnetic field. This manifests the presence of effective flux pinning centres in La doped compounds. As revealed by microstructural analyses the nanoparticles of $LaB_6$ (which are comparable in size to the coherence length) may be responsible for high intragrain $J_c$ values in the whole range of temperature and magnetic field investigated in the present work.

**Conclusion**

In conclusion, we have successfully synthesized La doped $MgB_2$ compounds by solid-state reaction at ambient pressure employing a special encapsulation technique. In the present investigation exploration of microstructural features induced by doping of La at Mg site of $MgB_2$ compound and its correlation with intragrain critical density ($J_c$) have been carried out. The highest $J_c$ value at 5K (~1.4 x $10^7$ A/cm$^2$ in self field, ~2.1 x $10^6$ A/cm$^2$, ~2.5 x $10^5$ A/cm$^2$ and ~1.8 x $10^4$ A/cm$^2$ at field of 1 T, 2.5T and 4.5T respectively), has been obtained for $Mg_{0.97}La_{0.03}B_2$ compound. This enhancement of $J_c$ for specific La doping ($Mg_{0.97}La_{0.03}B_2$) has been found to result due to high density of $LaB_6$ nanoparticle inclusions in $MgB_2$ grains which provide effective flux pinning centres.

**Acknowledgement**

The authors are grateful to Prof. A.R. Verma, Prof. C.N.R. Rao, Prof. A.V. Narlikar and Prof. T.V.Ramakrishnan for fruitful discussions. We put our sincere thanks to Dr. N.P. Lalla for his kind help in EDAX measurements. Financial supports from UGC and CSIR are gratefully acknowledged.

**Figure Captions**

Fig. 1:      Temperature dependence of dc magnetic susceptibility for $Mg_{1-x}La_xB_2$ (with x=0.01, 0.03, 0.05)

Fig. 2(a):      A representative TEM micrograph of $Mg_{0.99}La_{0.01}B_2$ compound. Presence of low density of $LaB_6$ nanoparticles (some of which are marked by $\rightarrow$) may be noticed.

Fig. 2(b):      Selected area diffraction pattern from the region shown in Fig. 2(a). The splitting of spots is due to the presence of low angle grain boundaries.

Fig. 2(c):      TEM micrograph corresponding to $Mg_{0.97}La_{0.03}B_2$ compound, showing the high density of $LaB_6$ nanoparticles (~5 to ~15 nm) within the $MgB_2$ grains.

Fig. 2(d):      Selected area diffraction pattern corresponding to micrograph Fig. 2(c) depicts spotty ring pattern which correspond to $MgB_2$ and $LaB_6$.

Fig. 2(e):      TEM micrograph corresponding to $Mg_{0.95}La_{0.05}B_2$ compound revealing $LaB_6$ nanoparticles which are bigger in size (~20 to ~35 nm) than those found for $Mg_{0.97}La_{0.03}B_2$.

Fig. 2(f):      Selected area diffraction pattern corresponding to micrograph Fig. 2(e) reveals hexagonal arrangement of diffraction spots of $MgB_2$ along with square diffraction spots of $LaB_6$ which has cubic lattice (marked by $\rightarrow$).

Fig. 3:      Intragrain critical current density (estimated on fine powder version of the samples) as a function of applied magnetic field at a temperature of 5K, 10K, 20K and 30K for (a) $MgB_2$ (b) $Mg_{0.99}La_{0.01}B_2$, (c) $Mg_{0.97}La_{0.03}B_2$, (d) $Mg_{0.95}La_{0.05}B_2$, superconductors. Highest critical current density for the $Mg_{0.97}La_{0.03}B_2$ compound may be noticed.

<u>Table-1</u>

Comparision of intragrain $J_c$ (A/cm$^2$) at 5K for undoped MgB$_2$ and optimally doped Mg$_{0.97}$La$_{0.03}$B$_2$

| Compositions | Intragrain $J_c$ (A/cm$^2$) at 5K | | | |
|---|---|---|---|---|
| | Self field | 1T | 2.5T | 4.5T |
| Undoped MgB$_2$ | $3.0 \times 10^5$ | $9.4 \times 10^4$ | $1.8 \times 10^4$ | $8.7 \times 10^2$ |
| Mg$_{0.97}$La$_{0.03}$B$_2$ (optimally doped) | $1.4 \times 10^7$ | $2.1 \times 10^6$ | $2.5 \times 10^5$ | $1.8 \times 10^4$ |

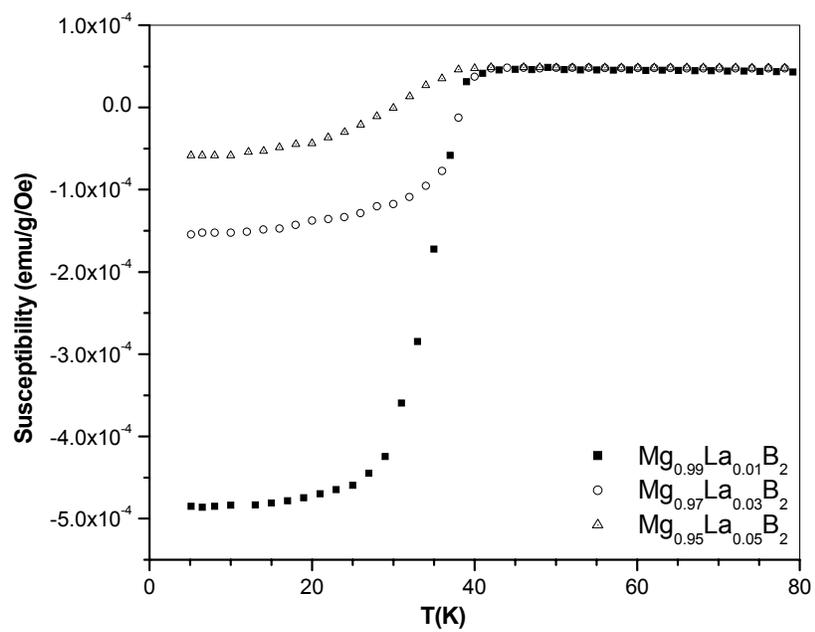

**Fig. 1**

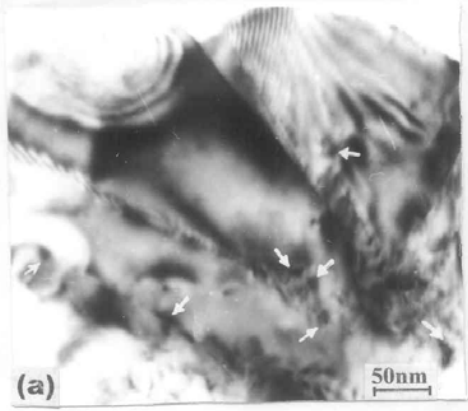

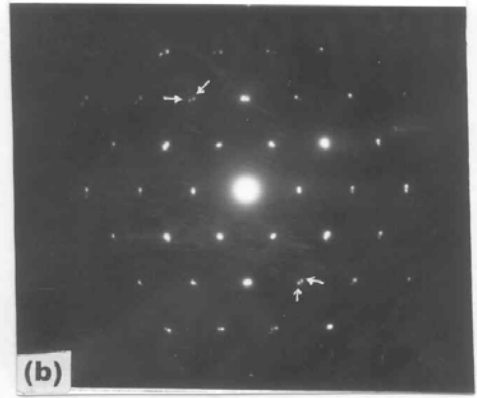

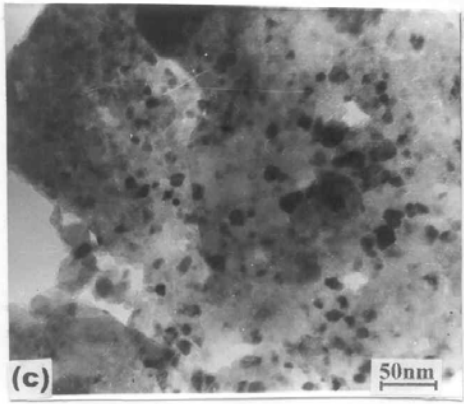

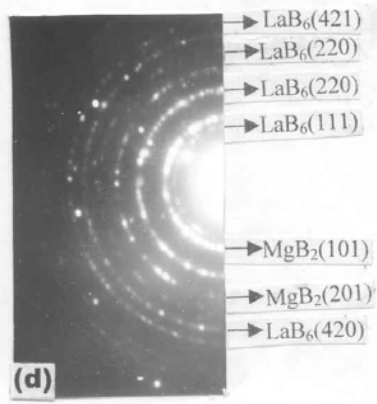

- LaB$_6$(421)
- LaB$_6$(220)
- LaB$_6$(220)
- LaB$_6$(111)
- MgB$_2$(101)
- MgB$_2$(201)
- LaB$_6$(420)

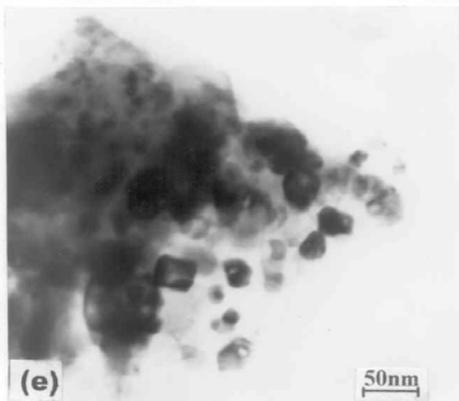

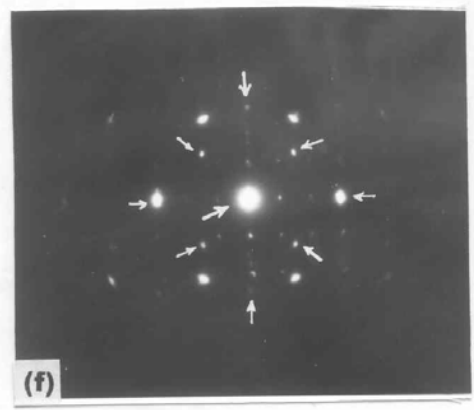

Fig.2

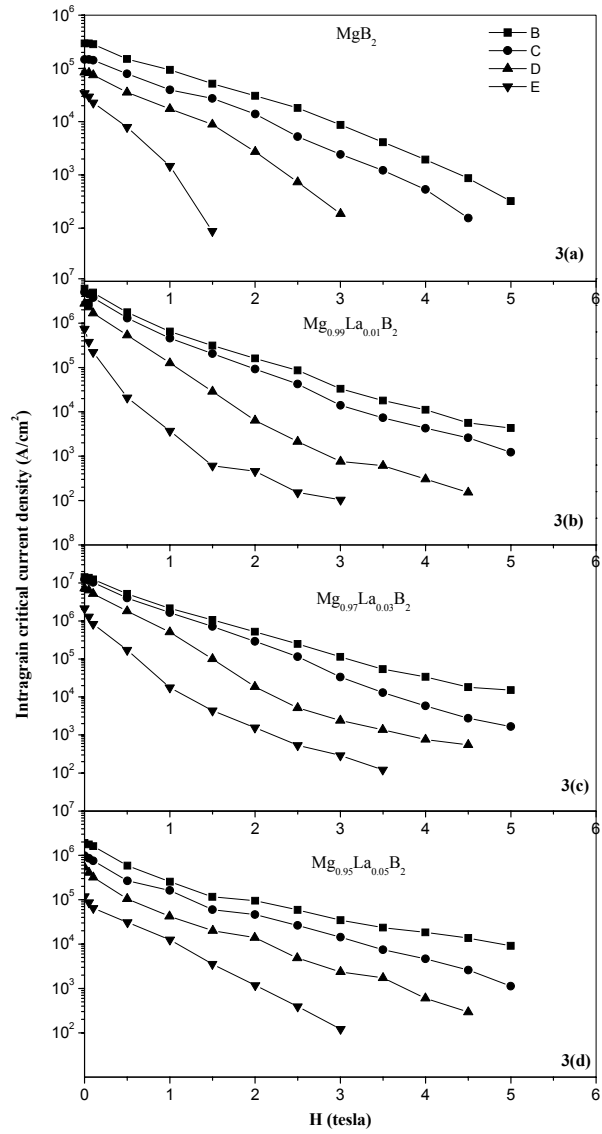

**Fig. 3**